\documentclass[conference,a4paper]{IEEEtran}

\usepackage{cite}

\usepackage[pdftex]{graphicx}
\graphicspath{{./graphics/}}

\usepackage[cmex10]{amsmath}
\usepackage{amssymb,amsfonts,relsize,bbm}		
\usepackage[utf8]{inputenc}
\usepackage[T1]{fontenc}
\usepackage[utf8]{inputenc}
\usepackage{tikz}
\usetikzlibrary{shapes,arrows,chains,arrows,decorations.pathmorphing,backgrounds,positioning,fit,petri,calc,patterns}
\usetikzlibrary{matrix,positioning}
\usepackage{pgfplots}
\usepackage{caption,subcaption}
\usepgfplotslibrary{groupplots}
\pgfplotsset{compat=newest}

\usepackage[pscoord]{eso-pic}

\usepackage{color}
\usepackage{algorithm,algorithmicx,algpseudocode}
\algdef{SE}[DOWHILE]{Do}{doWhile}{\algorithmicdo}[1]{\algorithmicwhile\ #1}		
\algrenewcommand\algorithmicforall{\textbf{foreach}}
\algrenewcommand\algorithmicrequire{\textbf{Input:}}
\algrenewcommand\algorithmicensure{\textbf{Output:}}

\newcommand{\rephrase}[1]{{\color{gray}#1}}
\newcommand{\gammaT}{\gamma^{\text{T}}}

\DeclareMathOperator{\E}{E}

\newcommand{\Pmax}{P^{\text{max}}}  
\newcommand{\acir}{\lambda}
\newcommand{\acirf}{\lambda_{|f'-f|}}

\newcommand{\Xb}{\mathbf{X}}
\newcommand{\xb}{\mathbf{x}}
\newcommand{\Yb}{\mathbf{Y}}
\newcommand{\Zb}{\mathbf{Z}}
\newcommand{\Pb}{\mathbf{P}}
\newcommand{\Hb}{\mathbf{H}}
\newcommand{\Ub}{\mathbf{U}}

\newcommand{\wb}{\mathbf{w}}
\newcommand{\Ns}{\mathcal{N}}
\newcommand{\Fs}{\mathcal{F}}
\newcommand{\Ts}{\mathcal{T}}
\newcommand{\Rs}{\mathcal{R}}

\newcommand{\Ss}{\mathcal{S}}
\newcommand{\Ls}{\mathcal{L}}
\newcommand{\Ps}{\mathcal{P}}
\newcommand{\Zs}{\mathcal{Z}}
\newcommand{\Qs}{\mathcal{Q}}

\newcommand{\js}{$j^{\mathrm{'s}}$}
\newcommand{\Xout}{X^{\mathrm{out}}}
\newcommand{\Xoutb}{\mathbf{X}^{\mathrm{out}}}
\newcommand{\Yout}{Y^{\mathrm{out}}}
\newcommand{\Youtb}{\mathbf{Y}^{\mathrm{out}}}
\newcommand{\Yoriginal}{Y^{\mathrm{original}}}
\newcommand{\davg}{$d_{\text{avg}}$}
\newcommand{\Ct}{C^{\text{T}}}
\newcommand{\figref}[1]{Fig.~\ref{#1}}
\newcommand{\comments}[1]{}
\IEEEoverridecommandlockouts

\usepackage{glossaries}
\makeglossaries
\newacronym{PA}{PA}{power amplifier}

\begin{document}
\title{Joint Scheduling and Power Control for V2V Broadcast Communication with Adjacent Channel Interference}
\author{Anver~Hisham,
	Di~Yuan,~\IEEEmembership{Senior~Member,~IEEE,}
	Erik~G.~Str\"{o}m,~\IEEEmembership{Senior~Member,~IEEE,} \\
	and~Fredrik~Br\"annstr\"om,~\IEEEmembership{Member,~IEEE}
	\thanks{Anver Hisham, Erik G. Str\"{o}m, and Fredrik Br\"annstr\"om are with the Division of Communication Systems, Department of Electrical Engineering, Chalmers University of Technology, SE-412 96 Gothenburg, Sweden. E-mail: \{anver, erik.strom, fredrik.brannstrom\}@chalmers.se}
	\thanks{Di Yuan is with the Department of Science and Technology, Link\"{o}ping University, 60174 Norrk\"{o}ping, Sweden. E-mail: di.yuan@liu.se}
		\thanks{The research was, in part, funded by the Swedish Governmental Agency for Innovation Systems (VINNOVA), FFI - Strategic Vehicle Research and Innovation, under Grant No. 2014-01387. This work has, in part, been performed in the framework of the H2020 project 5GCAR co-funded by the EU. The authors would like to acknowledge the contributions of their colleagues. The views expressed are those of the authors and do not necessarily represent the project. The consortium is not liable for any use that may be made of any of the information contained therein.}\\
	}

\maketitle	

\section{Introduction}
\label{sec:introduction}

\begin{abstract}
	This paper investigate how to mitigate the impact of adjacent channel interference (ACI) in vehicular broadcast communication, using scheduling and power control. Our objective is to maximize the number of connected vehicles. First, we formulate the joint scheduling and power control problem as a mixed Boolean linear programming (MBLP) problem. From this problem formulation, we derive scheduling alone problem as Boolean linear programming (BLP) problem, and power control alone problem as an MBLP problem. Due to the hardness in solving joint scheduling and power control for multiple timeslots, we propose a column generation method to reduce the computational complexity. We also observe that the problem is highly numerically sensitive due to the high dynamic range of channel parameters and adjacent channel interference ratio (ACIR) values. Therefore, we propose a novel sensitivity reduction technique, which can compute the optimal solution. Finally, we compare the results for optimal scheduling, near-optimal joint scheduling and power control schemes, and conclude that the effective scheduling and power control schemes indeed significantly improve the performance.
\end{abstract}

\subsection{Motivation}

\begin{figure}[t]	
	\centering	
	\vspace*{0.5cm}
\begin{tikzpicture}[rec/.style={rounded corners,inner sep=10pt,draw}]
\node(VUEi) at (0,0) [rec]{VUE $i$} ;
\node(VUEj) at (4,0) [rec]{VUE $j$} ;
\node(VUEk) at (7,0) [rec]{VUE $k$} ;
\draw[bend left,->,color=green] (VUEi) to node[above,black]{Desired link} (VUEj);
\draw[bend right,->,color=red,thick] (VUEk) to node[above,black]{Interference link} (VUEj);
\draw[bend left,->,color=green] (VUEi) to node[below,black]{$H_{i,j}$} (VUEj);
\end{tikzpicture}
	\caption{System model}\label{drawing:SystemModel}
	\vspace*{0.7cm}
\pgfdeclarelayer{bg}    
\pgfsetlayers{bg,main}
\tikzstyle{gArea} = [pattern=north west lines, draw=green,pattern color=green]
\tikzstyle{rArea} = [pattern=north east lines, draw=red,pattern color=red]
\tikzstyle{lStyle} = [dash dot,thick,black!50]
\begin{tikzpicture}
\begin{scope}[yshift=-4cm,scale = 1.2]{Receiver}
\draw[thick,->] (-.25,0) -- (7,0) node[anchor=north east]{Frequency};
\draw[thick,->] (0,-.25) -- (0,3.5);
\node[anchor=west,rotate=90] at (-.2cm,0.1) {Rx Power (dBm/Hz)};
\begin{pgfonlayer}{bg}
\draw[gArea] (.2,0) {[rounded corners] -- ++(30:1.5) -- ++(0.1,1) -- node(A){} ++(1,0) --++(0.1,-1) -- ++(-30:1.5)} -- cycle;
\draw[rArea,xshift=4cm] (-3.3,0)  {[rounded corners] -- ++(30:3.8) -- ++(0.2,1) -- node(B){} ++(1,0) --++(0.2,-1)}  {[opacity=0]  -- ++(-30:1) -- ++(0,-1.38)} -- cycle;
\end{pgfonlayer}
\draw[lStyle,dash phase=40pt] (B) --  ++(-3.6cm,0) node(Bl){};
\draw[lStyle] ($(A)+(4.5,-.8)$) node(Adr){}  -- ++(-5.5cm,0) node(Al){};
\draw[<->,thick,black!50] ($(Bl)+(0.2,-1pt)$) -- node[left,black]{ACIR} ++(0,-1.9);

\draw[lStyle,dash phase=13pt] ($(A)$)  -- ++(4.5cm,0) node(Ar){};
\draw[<->,thick,black!50] ($(Ar)+(-.1,-.05)$) -- node[right,black]{$\mathrm{SIR}_{i,j}$} ++(0,-.7);

\node[above,text width=2cm, align=center]  at ($(A)+(0,1.2)$) {Received signal from transmitter $i$};
\node[above,text width=2cm, align=center]  at ($(B)+(0,0)$) {Received signal from transmitter $k$};
\end{scope}
\end{tikzpicture}
	\caption{Received power spectral density at receiving VUE $j$.}\label{drawing:ACI}
\end{figure}

Vehicle-to-vehicle (V2V) communication can reduce traffic accidents significantly by broadcasting up-to-date local and emergency informations to nearby vehicles. To this end, both periodic and event-driven messages are conveyed. Periodic messages are transmitted by all vehicular user equipments (VUEs) in order to convey its current status to neighbors such as position, velocity and acceleration, whereas event-driven messages are sent when any emergency situation has been detected. However, conveying such safety related messages requires the establishment of highly reliable, low latency broadcast communication links between VUEs.

In a typical cellular communication systems, the reliability and the latency of a communication system is  limited significantly by co-channel interference (CCI), which is cross talk between transmitters scheduled in same frequency slot. 
However, in V2V communication with sufficiently dedicated frequency spectrum, CCI can be avoided by scheduling VUEs in non-overlapping frequency slots. But scheduling two VUEs in nearby frequency slots result in adjacent channel interference (ACI), which is the interference due to the spillage of transmit power to nearby frequency slots than the intended frequency slot. ACI is mainly due to nonlinearity of the \gls{PA} in transmitter.  Advanced methods have been developed to linearize \gls{PA} \cite{Mohammadi1,Lavrador1,Kenington1,Jessica1}, however, the clipping effect of \gls{PA} cannot be avoided, which results in ACI. An example of ACI is illustrated in \figref{drawing:SystemModel}--\ref{drawing:ACI}, where VUE $i$ is transmitting to VUE $j$, and VUE $k$ is interfering. We show that signal to interference ratio ($\mathrm{SIR}_{i,j}$) of VUE \js~reception is affected by the ACI from VUE $k$, even though VUE $k$ is transmitting on an adjacent frequency slot.

A parameter named adjacent channel interference ratio (ACIR) is widely used to measure the ACI in neighboring frequency slots. ACIR is defined as the ratio of received signal power in the transmitted frequency slot to the received ACI in the nearby frequency slot. In other words, ACIR is the ratio between the average in-band received power from transmitter $k$ to the average out of band received power from transmitter $k$'s signal in the frequency slot allocated for transmitter $i$, as illustrated in \figref{drawing:ACI}.

\subsection{State of the Art}
Typical cellular communication is limited by CCI, due to the spectral re-usage. Therefore, most of the existing literature consider approaches to mitigate CCI alone \cite{v2vsch1,v2vsch2,v2vsch3}. However, in a V2V communication scenario with dedicated spectrum, we can minimize CCI by allocating maximum VUEs in nonoverlapping frequency slots. But in the absence of CCI, the communication link performance is majorly limited by ACI \cite{AnverICC}. \comments{In \cite{Albasry1}, the authors analyze the impact of ACI for device-to-device (D2D) communication, for various user densities and transmit power, and conclude that ACI indeed causes outage in the system when the user density is high. Similar analysis has been done in \cite{Li1}, for the impact of ACI from cellular uplink to D2D communication, and the authors conclude that ACI indeed has a large negative impact on D2D communication. In \cite{Haas1}, the authors analyzed the probability density function (PDF) of ACI for a cellular system based upon the Universal Mobile Telecommunications System (UMTS).} Extensive studies have been done to measure the impact of ACI when different communication technologies coexist in adjacent frequency bands \cite{acib1,acib2,acib3,acib4}. In \cite{Xia1}, the authors assess the performance degradation due to ACI when two LTE base stations are deployed in adjacent frequency channels. Studies have been done to measure the impact of ACI when different communication technologies coexist in adjacent frequency bands \cite{acib1,acib2,acib3,acib4}, and the impact of ACI on 802.11b/g/n/ac was also broadly studied \cite{aciw1,aciw2,aciw3}. However, adequate attention has not yet been given to study the effect of ACI within a V2V broadcast communication scenario. To further understand the impact of ACI, readers are directed to our previous works \cite{AnverICC,AnverArchive}.

\subsection{Contributions}
Our goal is to maximize the number of connected VUEs in a vehicular network, using proper scheduling and power control schemes. We make following contributions in this paper;
\begin{enumerate}
	\item We formulated joint scheduling and power control problem in order to maximize the number of connected VUEs, in the presence of both CCI and ACI, as a mixed Boolean linear programming (MBLP) problem.
	\item We formulated scheduling for a fixed power as a Boolean linear programming (BLP) problem, and power control for a fixed schedule as an MBLP problem.
	\item Due to the high computational complexity of the joint scheduling and power control problem, we propose a novel column generation method which can linearize the computational complexity with respect to the number of timeslots $T$.
	\item The scheduling problem is highly sensitive due to the high dynamic range of channel parameters and ACIR values. Thus, computing the optimal schedule is extremely hard. Therefore, we propose a new method to reduce the sensitivity of the computation of the optimal schedule, inspired from \cite{Capone1}.
\end{enumerate}

\section{Preliminaries}
\label{sec:preliminaries}

\subsection{Notation} 
We use the following notation throughout the paper. Sets are denoted by calligraphic letters, e.g., $\mathcal{X}$, with $|\mathcal{X}|$ denoting its cardinality, and $\emptyset$ indicating an empty set. Lowercase and uppercase letters, e.g., $x$ and $X$, represent scalars. Lowercase boldface letters, e.g., $\xb$, represent a vector where $x_{i}$ is the $i^{\mathrm{th}}$ element and $|\xb|$ is its dimensionality. The uppercase boldface letters, e.g., $\mathbf{X}$, denote matrices where $X_{i,j}$ indicates the $(i,j)^\mathrm{th}$ element. 
The notation $\mathbbm{1}\{\textit{statement}\}$ is either 1 or 0, depending upon if the \textit{statement} is true or false. 

\subsection{Assumptions}
We have following assumptions;
\begin{enumerate}
	\item We define $\Ns = \{1,2,\cdots,N\}$ as the set of VUEs,  $\Fs = \{1,2,\cdots,F\}$ as the set of frequency slots, and $\Ts = \{1,2,\cdots,T\}$ as the set of timeslots for scheduling. A resource block (RB) is a frequency slot in a timeslot, that is, the frequency slot $f$ in the timeslot $t$ is denoted by RB $(f,t)$. 
	\item A VUE $i \in \Ns$ want to broadcast its packet to the VUEs in the set $\Rs_i \subset \Ns$. For convenience, we define the set of intended transmitters for receiver VUE~$j$ as $\mathcal{T}_j  \triangleq \{i: j \in \mathcal{R}_i\}$. We note that $\mathcal{T}_j$, $j\in\Ns$ is completely determined by $\mathcal{R}_i$, $i\in\Ns$ and vice versa. As an example, the set $\mathcal{R}_i$ could be all vehicles within a certain distance from VUE~$i$; however, the proposed method does not rely on any particular structure for $\mathcal{R}_i$ or, therefore, $\mathcal{T}_j$. Moreover, we define the set $\Ls = \{(i,j): i \in \Ns, j \in \Ts_i  \}$ as the set of all intended links.
	\item A centralized controller exists which can schedule and power control all VUEs in the network in $\Fs \times \Ts$ RBs. Large-scale channel parameters (i.e., pathloss and penetration loss) are assumed to be slowly varying compared to the scheduling interval $T$. Therefore, we assume that the centralized controller has access to the slowly varying channel state information (CSI) between all pairs of VUEs to compute the average SINR. A base station (BS) or a VUE can act as a centralized controller.
	\item  The maximum transmit power of a VUE is limited to $\Pmax$.
	\item A VUEs can successfully transmit a message in an RB, if the received SINR is above a certain threshold $\gammaT$ \cite[Lemma 1]{Wanlu2016} .
\end{enumerate}

\subsection{System Model} 

Key mathematical symbols are listed in Table \ref{table_notation}. We indicate a transmitting VUE as VUE $i$, receiving VUE as VUE $j$, and interfering VUE as VUE $k$, as illustrated in \figref{drawing:SystemModel}. Similarly the link $(i,j)$ indicate the link from VUE $i$ to VUE $j$. The parameter $H_{i,j}$ is the average channel power gain from VUE $i$ to VUE $j$. Hence, $H_{i,j}$ takes into account pathloss and large-scale fading between VUE $i$ and VUE $j$. 

Assume that VUE $i$ is transmitting in RB $(f,t)$ and VUE $k$ in RB $(f',t)$.  If interferer VUE $k$ is transmitting on the same RB as VUE $i$ (i.e., $f'=f$), then VUE \js~reception is affected by CCI from VUE $k$. On the other hand, if VUE $k$ is transmitting on a nearby frequency slot of VUE $i$  (i.e., $f' \neq f$), then the reception is affected by ACI instead. In this paper, we consider overlapping scheduling, i.e., multiple VUE can be scheduled in the same RB.


\section{Problem Formulation} \label{sec:joint}

\subsection{Joint Scheduling and Power Control Problem}
Let $\mathbf{X} \in \{0,1\}^{N \times F \times T }$ be the scheduling matrix defined as follows,
\begin{equation}
X_{i,f,t} \triangleq \left\{ \begin{array}{rll}
1, & \parbox{5cm}{\raggedright if VUE $i$ is scheduled in RB $(f,t)$ } \\
0, & \parbox{5cm}{\raggedright otherwise.} \\
\end{array} \right. 
\end{equation}

 Similarly, $P_{i,f,t}$ is the transmit power of VUE $i$ in RB $(f,t)$. The variable $P_{i,f,t}$ is constrained by the maximum transmit power $\Pmax$ of a VUE in a timeslot as follows,
  \begin{equation}
 \sum_{f = 1}^{F} P_{i,f,t} \leq \Pmax  \qquad \qquad \, \forall\, i,\, t  
 \end{equation}
 Moreover, $P_{i,f,t}$ is also constrained by scheduling as follows,
 \begin{equation}
 0 \leq P_{i,f,t} \leq \Pmax X_{i,f,t} \qquad \qquad \forall i,\, f,\, t  
 \end{equation}

Let us consider a link $(i,j)$ in RB $(f,t)$, i.e., the link from VUE $i$ to VUE $j$ in frequency slot $f$ and timeslot $t$. The total signal power $S_{i,j,f,t}$ and interference power $I_{i,j,f,t}$ received by VUE $j$ while decoding the signal from VUE $i$ in RB $(f,t)$ can be computed as follows,

\begin{align}
S_{i,j,f,t} &= P_{i,f,t} H_{i,j} \label{definitionS}~,   \\
I_{i,j,f,t} &= \overset{F}{\underset{f'=1}{\sum}}  \overset{N}{\underset{\substack{k = 1 \\ k \neq i}}{\sum}}  \acirf  P_{k,f',t} H_{k,j}~, \label{definitionI}   
\end{align}
where $\acir_r$ is the adjacent channel interference ratio (ACIR) from a frequency slot $f$ to frequency slot $f \pm r$. Therefore, $\acirf$ is the ACIR from frequency slot $f'$ to $f$. Note that when $f'=f$, then the interference is CCI, instead of ACI. Therefore, in order to accommodate CCI, we make $\acir_{0}=1$. 

Following (\ref{definitionS})  and (\ref{definitionI}), we can compute SINR $\Gamma_{i,j,f,t}$ of the link $(i,j)$ in RB $(f,t)$, as follows,
\begin{equation}
\hspace{-2em}\Gamma_{i,j,f,t} = \frac{ S_{i,j,f,t}} {\sigma^2 +   I_{i,j,f,t} },
\end{equation}
where $\sigma^2$ is the noise power in an RB. 

For the link to be successful, SINR must be above a certain threshold $\gammaT$, i.e.,
\begin{align}
\Gamma_{i,j,f,t} &\geq \gammaT  \\
\Leftrightarrow \quad S_{i,j,f,t} - \gamma^{\mathrm{T}} I_{i,j,f,t} &\geq \gamma^{\mathrm{T}} \sigma^2.   \label{constraintSingleLink1}
\end{align}
However, it might not be possible to fulfill this condition for all links $(i,j)$ in all RBs $(f, t)$. To select which combinations of $i,~j,~f$, and $t$ to enforce this condition, we introduce the matrix $\mathbf{Y}\in\{0,1\}^{N\times N\times F\times T}$, where
\begin{equation}
\label{Ydefinition}
Y_{i,j,f,t} \triangleq
\begin{cases}
1, & \text{if \eqref{constraintSingleLink1} is enforced}\\
0, & \text{otherwise}
\end{cases}
\end{equation}

\begin{table}[t] 
	\caption{Key Mathematical Symbols}\label{table_notation}
	\renewcommand{\arraystretch}{1.3}
	\begin{tabular}[t]{|cl|}
		\hline		Symbol  &   Definition  \\
		$N$ & Number of VUEs \\
		$F$ & Number of frequency slots \\
		$T$ & Number of timeslots \\
		$\gammaT$ & \parbox[t]{7cm}{SINR threshold to declare a link as successful}\\
		$\sigma^2$ & Noise power in an RB \\		
		$\Pmax$ & Maximum transmit power of a VUE \\
		$P_{i,f,t}$ & Transmit power of VUE $i$ in an RB in timeslot $t$ \\
		$H_{i,j}$ & Average channel power gain from VUE $i$  to VUE $j$ \\
		$\acir_r$ & ACI from any frequency slot $f$ to frequency slot $f \pm r$ \\ 
		$X_{i,f,t}$ &  Indicate if VUE $i$ is scheduled to transmit in RB $(f,t)$ \\
		$Y_{i,j,f,t}$ & Indicate if link $(i,j)$ is successful in RB $(f,t)$ \\
		$\Gamma_{i,j,f,t}$ & SINR of the link $(i,j)$ in RB $(f,t)$\\
		$V_{i,k,r,t}$ &  \parbox{7cm}{Indicate if VUE $i$ and $k$ are scheduled not more than \\ $r$ frequency slots apart in timeslot $t$}\\		
		\hline
	\end{tabular}
\end{table}

We can combine~\eqref{constraintSingleLink1} and~\eqref{Ydefinition} into a single constraint,
\begin{equation}
\label{constraintMatrix2}
S_{i,j,f,t} - \gamma^{\mathrm{T}} I_{i,j,f,t} \ge \gamma^{\mathrm{T}}
\sigma^2 - \eta(1-Y_{i,j,f,t})		
\end{equation}
where $\eta$ is a sufficiently large number to make~\eqref{constraintMatrix2} hold whenever $Y_{i,j,f,t}=0$, regardless of the schedule and power allocation. It is not hard to show that $\eta = \gamma^{\mathrm{T}}(NP^{\mathrm{max}}+\sigma^2)$ is sufficient.

Observe that if $Y_{i,j,f,t} = 1$, then the link $(i,j)$ is successful in RB $(f,t)$ as per~\eqref{constraintMatrix2}. Similarly, let $Z_{i,j}$ indicate the success of the link $(i,j)$ in any RB $(f,t)$, where $f \in \Fs$ and $t \in \Ts$, i.e.,
\begin{subequations}
\begin{align}
Z_{i,j} &\triangleq
\begin{cases}
1, & \text{if link $(i,j)$ is successful,}   \label{ZijDefinition} \\
0, & \text{otherwise,}
\end{cases}  \\
&= \min\{1,   \sum_{t=1}^T\sum_{f=1}^F Y_{j,j,f,t}\}~,  \label{ZijComputation}  
\end{align}
\end{subequations}
where the minimum in~\eqref{ZijComputation} is required in order to not to count successful links between VUE~$i$ and VUE~$j$ more than once.

We can translate (\ref{ZijComputation}) into the following set of linear constraints,
\begin{subequations}
\begin{align}
	Z_{i,j} &\geq Y_{i,j,f,t}  \qquad\qquad \forall\,f,\,t   \label{Zij_lowerbound} \\
	Z_{i,j} &\leq \sum_{t=1}^{T} \sum_{f=1}^{F}   Y_{i,j,f,t}  \quad \forall\,f,\,t  \\
	Z_{i,j} &\in \{0,1\} \label{ZijijBooleanconstr}
\end{align}
\end{subequations}
We note that the constraint (\ref{Zij_lowerbound}) is redundant, since we do not require a lower bound for $Z_{i,j}$ while maximizing $\sum\limits_{(i,j) \in \Ls}   Z_{i,j} $. Additionally, the last boolean constraint (\ref{ZijijBooleanconstr}) can be relaxed to the constraint $Z_{i,j} \leq 1$, since we are trying to maximize $Z_{i,j}$, and $\Yb$ is a Boolean matrix.

Putting everything together, we arrive at the following MBLP problem,
\begin{subequations} \label{jointSCPC}
	\begin{align}
	& \max\limits_{\mathlarger{\mathbf{P},\Xb,\mathbf{Y},\mathbf{Z}}}  \sum\limits_{(i,j) \in \Ls}  Z_{i,j}  \label{jointSCPC_obj} \\
	&\mbox{s.t. } \nonumber \\
	&  P_{i,f,t} H_{i,j}   - \gammaT  \overset{N}{\underset{\substack{k = 1 \\ k \neq i}}{\sum}}  \overset{F}{\underset{f'=1}{\sum}}  \acirf P_{k,f',t} H_{k,j} \hspace*{1.5cm}    \nonumber \\
	& \hspace*{0.2cm}\geq \gammaT \sigma^2 - \gammaT(N\Pmax+\sigma^2) (1-Y_{j,j,f,t}) \hspace*{0.3cm} \forall \, i,j, f,t   \label{jointSCPC_sinr} \\[5pt] 
	&\sum_{f = 1}^{F} P_{i,f,t} \leq \Pmax  \hspace*{4.3cm}   \forall\, i,\, t \\
	& 0 \leq P_{i,f,t} \leq \Pmax X_{i,f,t}   \hspace*{3.2cm}  \forall\, i,f,t  \label{jointSCPC_power} \\				
	& Z_{i,j} \leq \sum_{t=1}^{T} \sum_{f=1}^{F}   Y_{i,j,f,t}  \hspace*{3.7cm} \forall\,i,\,j  \\
	& Z_{i,j} \leq 1	 \hspace*{5.7cm}  \forall\,i,j \\
	&\mathbf{X} \in \{0,1\}^{N \times F \times T }	\label{blp_booleanConstr}  \hspace*{2.8cm} 	 \\
	&\mathbf{Y} \in \{0,1\}^{N \times N \times F \times T }		\\
	&\Pb\in\mathbb{R}^{N \times F \times T} 
	\end{align}
\end{subequations}

Here are some key observations about the above problem formulation;
\begin{enumerate}
	\item The problem formulation (\ref{jointSCPC}) is for overlapping scheduling and power control. We observe that the Boolean variable $\Xb$ is unnecessary, and constraint (\ref{jointSCPC_power}) can be replaced with constraint $0 \leq P_{i,f,t} \leq \Pmax $. Here, we are doing only the power control, and we schedule VUEs in RB $(f,t)$ whenever its power in the RB is nonzero, i.e., $X_{i,f,t} = \mathbbm{1}\{P_{i,f,t} > 0\}$. This way we can reduce the computational complexity of the problem.
	
	\item The problem formulation (\ref{jointSCPC}) can be made into a nonoverlapping scheduling and power control problem (hence avoiding CCI), by adding an extra constraint as follows.
	\begin{equation}
	\sum_{i=1}^{N} X_{i,f,t}  \leq 1  \qquad  \forall\,f,\,t
	\end{equation}
	\item A VUE scheduling can be limited to maximum one RB in a timeslot using the following constraint,
		\begin{equation}
		\sum\limits_{f=1}^{F}  X_{i,f,t}	\leq 1 \quad\forall\,i,\,t    \label{constr_maxOneRB}
		\end{equation}	
		Limiting scheduling to maximum one RB in a timeslot reduces the computational complexity without much compromise on the performance, as we will see in Section \ref{sec:sensitivity}.
	
	\item The problem formulation (\ref{jointSCPC}) is for full-duplex communication, where a VUE can both transmit and receive simultaneously. However, (\ref{jointSCPC}) can be made into a half-duplex communication problem by adding the following extra constraint,
	\begin{equation}
	Y_{i,j,f,t} \leq (1-X_{j,f',t}) \quad \forall\,i\,j\,f\,f'\,t
	\end{equation}
	
	\item \label{SchedulingProblem} The problem formulation (\ref{jointSCPC}) can be translated into a  scheduling alone problem by changing (\ref{jointSCPC_power}) into the following constraint,
	\begin{equation}
	P_{i,f,t} = \bar{P}_{i,t} X_{i,f,t} \qquad\qquad  \forall\,i,f,t
	\end{equation}
	where $\bar{P}_{i,t}$ is the transmit power of VUE $i$ in timeslot $t$, which is a known value.	The resulting problem is a Boolean linear programming (BLP) problem.
	\item The problem formulation (\ref{jointSCPC}) can be translated into a power control alone problem by fixing scheduling (i.e., $X_{i,f,t}$), and making power values (i.e., $P_{i,f,t}$) as optimization variables. Once we know the scheduling, we do not need the constraint (\ref{jointSCPC_sinr}) for all VUEs and for all RBs (i.e., $\forall~i,j,f,t$), instead, we can limit this constraint only for the scheduled RBs for each VUE. The resulting problem is an MBLP problem.
	\item  The problem formulation (\ref{jointSCPC}) can be translated into a problem to maximize the minimum number of successful links for a VUE, instead of total number of successful links. In this way, we are guaranteeing atleast $L^*$ successful links for any VUE. This is done by changing the objective function (\ref{jointSCPC_obj}) as follows,
	\begin{equation}
	L^* = \max\limits_{\mathlarger{\mathbf{P},\Xb,\mathbf{Y},\mathbf{Z}}}    L
	\end{equation} 
	and adding an extra constraint to (\ref{jointSCPC}) as follows,
	\begin{equation}
	\sum\limits_{j \in \Rs_i}  Z_{i,j} \geq L  \qquad \forall\,i
	\end{equation}
	
\end{enumerate}

\section{Joint Scheduling And Power Control Using Column Generation Method}  
\label{sec:columnGeneration}

We observe that the worst-case computational complexity of (\ref{jointSCPC}) increases exponentially with respect to the number of Boolean variables, like a typical NP-hard BLP problem\cite{Papadimitriou}. Since there are $(N+1)NFT$ Boolean variables in our problem formulation in \eqref{jointSCPC}, we see that the worst-case computational complexity is $\mathcal{O}(2^{(N+1)NFT})$. Therefore, in this section, we propose an efficient approximation method to reduce the computational complexity from $\mathcal{O}(2^{(N+1)NFT})$ to $\mathcal{O}(T2^{(N+1)NF})$.

First, we explain the basic idea behind the algorithm. Let us assume that we have an ordered set of all possible power value matrices $\tilde{\Ps} = \{\tilde{\Pb}^1,\tilde{\Pb}^2,\cdots,\tilde{\Pb}^Q\}, ~ \tilde{\Pb}^q \in \mathbb{R}^{N \times F \times T}, ~\forall\, 1 \leq q \leq Q$, and the corresponding set of successful link matrices $\tilde{\Zs} = \{\tilde{\Zb}^1,\tilde{\Zb}^2,\cdots,\tilde{\Zb}^Q\}, ~ \tilde{\Zb}^q \in \{0,1\}^{N \times N}, ~\forall\, 1 \leq q \leq Q$. \comments{Ofcourse, this ordered set is unbounded ($|\tilde{\Ps}| = |\tilde{\Zs}| = \infty$).} Each element $\tilde{\Pb}^q \in \tilde{\Ps}, 1 \leq q \leq Q$ is a solution matrix $\Pb$ of the problem (\ref{jointSCPC}) for single timeslot (i.e., $T=1$), and the corresponding element $\tilde{\Zb}^q$ is the respective solution matrix $\Zb$. 
In other words, $\tilde{Z}^q_{i,j}$ indicate the success status of the link $(i,j)$, when we use the power values $\tilde{P}_{i,f,t}^q,~\forall\,i,f,t$.
Once we have the sets $\tilde{\Ps}$ and $\tilde{\Zs}$, our joint scheduling and power control problem is reduced to finding out the best $T$ elements out of the set $\tilde{\Zs}$, which would maximize the total number of successful links. This problem is stated as follows,
\begin{subequations} \label{masterProblem}
	\begin{align}
	c_Q &= \max\limits_{\Zb',\wb}  \sum\limits_{(i,j) \in \Ls}  Z'_{i,j}   \\
	\mbox{s.t. }& \nonumber \\
	&\sum\limits_{q=1}^{Q} \tilde{Z}_{i,j}^q w_q \geq Z'_{i,j} 	\quad\forall\,i,j \label{ZbarConstraint1}\\
	&\sum\limits_{q=1}^{Q} w_q \leq T	\label{wqConstraint1}\\
	&Z'_{i,j} \in \{0,1\}  \quad \forall\,i,\,j	\\
	& w_q \in \{0,1\}	\quad \forall\,q
	\end{align}
\end{subequations}
The Boolean vector $\wb \in \{0,1\}^Q$ indicates which of the $T$ elements from the set $\tilde{\Ps}$ are chosen.

However, there are two practical difficulties with this approach; 1) The set of all possible power matrices has got infinite cardinality (i.e., $|\tilde{\Ps}| = \infty$), 2) Even if $\tilde{\Ps}$ has finite cardinality (i.e., $|\tilde{\Ps}| < \infty$), then the problem formulation (\ref{masterProblem}) is still an NP-hard problem, since it is equivalent to a maximum coverage problem \cite{Vazirani1}. To overcome these practical difficulties, we propose a column generation method, in which we split the problem into two separate problems as 1) master problem and 2) subproblem, and solve those iteratively.
\begin{enumerate}
	\item \textit{Master problem}: We modify the problem formulation (\ref{masterProblem}) by relaxing the Boolean constraints (\ref{masterProblem}d--e) for $\Zb'$ and $\wb$, to the constraints $0\leq \Zb' \leq 1,~ 0\leq \wb \leq 1$. Let us call this new relaxed problem formulation as [M1]. The problem [M1] is easy to solve for a finite $Q$, since it is a linear programming (LP) problem.
	\item \textit{Subproblem}: We will not generate the set of all possible power vectors $\tilde{\Ps}$. Instead, we initialize $\tilde{\Ps} = \emptyset$ and iteratively add the power vectors to $\tilde{\Ps}$, and corresponding success status vector to $\tilde{\Zs}$ using the column generation method.
\end{enumerate}

The algorithm is an iterative algorithm, as explained in Algorithm \ref{Alg_columnGenerationMethod}. In each iteration, we solve the master problem [M1] and the subproblem. The master problem generates dual values $\Pi_{i,j}$ for each of the constraints \eqref{ZbarConstraint1}, and $\pi$ for the constraint \eqref{wqConstraint1}. The subproblem is same as the problem formulation (\ref{jointSCPC}) for a single timeslot (i.e., $T=1$), but with the modified objective function (i.e., $\max~\sum\limits_{(i,j) \in \Ls} \Pi_{i,j} Z_{i,j} - \pi $). In each iteration, the master problem passes the weights $\Pi$ and $\pi$ to the subproblem. The subproblem solves \eqref{jointSCPC} with the modified objective function, and generates an optimal $\Pb$ and $\Zb$ for a single timeslot. These solutions $\Pb$ and $\Zb$ from the subproblem are then added to the sets $\tilde{\Ps}$ and $\tilde{\Zs}$ respectively. In the subsequent iteration, the master problem compute the dual values using the augmented sets $\tilde{\Ps}$ and $\tilde{\Zs}$. 

Intuitively, in an iteration in master problem, the weights $\Pi_{i,j}$ would be larger for a recurrently failing link $(i,j) \in \Ls$ in the previous iterations, therefore, the subproblem would prioritize those links in the subsequent iterations. We stop the iterations, when the subproblem objective value is zero or negative, which indicate that the master problem cannot improve the solution anymore by augmenting the set $\tilde{\Pb}$ (i.e., by adding any extra elements to $\tilde{\Pb}$). After all the iterations, we have to choose best $T$ power vectors from the set $\tilde{\Pb}$. Since this problem is an NP-hard problem, we use an approximation algorithm here. We choose the power vectors corresponding to the $T$ highest values of $\wb$ as shown in Algorithm \ref{Alg_columnGenerationMethod}. \comments{, and the simulation results shows that this method provides close to optimal performance.}

Once we know the power value matrix $\Pb$, then we can compute the scheduling matrix $\Xb$ as $X_{i,f,t} = \mathbbm{1}\{P_{i,f,t} > 0\}\quad\forall\,i,f,t$.

Typical column generation method provides optimal solution if both master problem and slave problem are LP problems. Hence, the columns generation method presented in this section is an approximation method providing suboptimal solution, since 1) master problem \eqref{masterProblem} is solved using approximation method, 2) the slave problem is not an LP problem due to the presence of Boolean variable $\Yb$.


\begin{algorithm}[!t]
	\caption{Column Generation Method for Joint Scheduling and Power Control}
	\label{Alg_columnGenerationMethod}
	\begin{algorithmic}[1]
		\Require{$\{N,F,T,\Hb,\mathbf{\acir},\tilde{\Pb},\gammaT,\sigma^2 \}$}
		\Ensure{$\Pb$}  
		\Statex // Compute the set of power vectors $\{\tilde{\Pb}^1,\tilde{\Pb}^2,\cdots,\tilde{\Pb}^Q \}$   
		\State Initialize $Q=0,~ \tilde{\Pb}^1 = 0^{N \times F \times T},~ \tilde{\Zb} = 0^{N \times N}$.
		\Do 
		\State $Q = Q +1$
		\State Solve the master problem, i.e., [M1]. Store the optimum objective value as $c_Q$. Compute the dual values $\Pi_{i,j}$ and $\pi$ for the constraints in (\ref{masterProblem}b--c).
		\State Solve the subproblem, i.e., problem formulation (\ref{jointSCPC}) after modifying the objective function \eqref{jointSCPC_obj} to $\max \sum_i \sum_i \Pi_{i,j} Z_{i,j} - \pi $, for a single timeslot. Compute the solutions $\Zb$ and $\Pb$, and assign $\tilde{\Zb}^{Q+1} = \Zb, \tilde{\Pb}^{Q+1} = \Pb$.
		\doWhile{\rephrase{objective value of the subproblem is positive}}
		\Statex // Select $T$ power vectors from $\tilde{\Ps}$
		\For{t=1:T}
			\State $q^* = \underset{1 \leq q \leq Q}{\arg\max}  \{w_q\}$
			\State Assign the power value matrix $\tilde{\Pb}^{q^*}$ to the timeslot $t$, i.e., $P_{i,f,t} = \tilde{P}_{i,f,t}^{q^*} ~\forall\,i,f$.
			\State $w_{q^*}=0$
		\EndFor 
		\State $X_{i,f,t} = \mathbbm{1}\{P_{i,f,t} > 0\}  \qquad  \forall\,i,f,t$
	\end{algorithmic}
\end{algorithm}

\section{Methods to Reduce Sensitivity of (\ref{jointSCPC_sinr}) for Scheduling Problem}   \label{sec:sensitivity}
We note that, the high sensitivity of the constraint (\ref{jointSCPC_sinr}) due to the presence of both large and small coefficients, makes the problem (\ref{jointSCPC}) harder to solve. To overcome this sensitivity issue, we apply a novel computational approach inspired from \cite{Capone1}, by replacing  (\ref{jointSCPC_sinr}) with more tractable Boolean cover inequalities. In this section, we assume a VUE is scheduled to maximum one RB in a timeslot. This is a reasonable assumption without compromising the performance quality, since scheduling in multiple RBs in a timeslot results in lesser transmit power available for transmission in each RB, and disseminate the interference over multiple RBs.

\subsection{Preliminaries}
In this subsection, we define the variables required for the implementation of the sensitivity reduction method. Let $\bar{I}_{i,j}$ denote the maximum tolerable interference power for a link $(i,j)$. That is, if the received interference power is less than or equals to $\bar{I}_{i,j}$, then the link $(i,j)$ is successful, otherwise, the link is a failure. The value of $\bar{I}_{i,j}$ can be computed from (\ref{constraintSingleLink1}) as follows,
\begin{equation}
\bar{I}_{i,j} = \frac{\bar{P}_{i,t}H_{i,j}-\gammaT \sigma^2}{\gammaT}  \label{Iij_computation}
\end{equation}
where $\bar{P}_{i,t}$ is the transmit power of VUE $i$ if scheduled in timeslot $t$. We assume $\bar{P}_{i,t}$ is a known quantity since we are considering only scheduling here. A zero value, i.e., $\bar{P}_{i,t}=0$, indicate that VUE $i$ cannot be scheduled in timeslot $t$. 

Let us introduce a new variable $V_{i,k,r,t} \in \{0,1\}$ indicating if VUE $i$ and $k$ are scheduled in RBs which are not more than $r$ RBs far apart in timeslot $t$. That is,
\begin{equation} \label{define:Vikrt}
V_{i,k,r,t}  = 
\begin{cases}
1, &  \parbox{5cm}{ $\exists (f,f'), \text{  s.t. }  X_{i,f,t}=1, \\ \hspace*{0.2cm} X_{k,f',t}=1, |f'-f| \leq r,$}    \\
0, & \text{Otherwise}
\end{cases}
\end{equation}
The variable $V_{i,k,r,t}$ can be computed linearly using the variables $\Xb$ and $\Yb$ as explained in Appendix \ref{Appendix:V_definition}.

\subsection{Algorithm Description}

First, we modify the joint scheduling and power control problem formulation in (\ref{jointSCPC}) into a scheduling alone problem using the technique explained in \ref{SchedulingProblem}) in Section \ref{sec:joint}. We then remove the constraint (\ref{jointSCPC_sinr}) from the problem formulation (\ref{jointSCPC}), which is the only sensitive constraint in (\ref{jointSCPC}). We call this modified problem as [M2].

The procedure here is an iterative cutting plane method as shown in Algorithm \ref{Alg_sensitiviyRemoval}, where in each iteration we solve [M2] by adding extra cover inequalities. In an iteration, we solve the problem formulation [M2]  and call the resulting solution scheduling matrix as $\Xoutb$ and successful link matrix as $\Youtb$. Let the Boolean variable $\Yoriginal_{i,j,f,t}$ indicate if the link $(i,j)$ is originally successful or not. Given the scheduling matrix $\Xoutb$, we can compute $\Yoriginal_{i,j,f,t}$ as follows,
\begin{equation}
\Yoriginal_{i,j,f,t} = \mathbbm{1} \{ \frac{\Xout_{i,f,t} \bar{P}_{i,t} H_{i,j}}{ \sigma^2 + \sum\limits_{\substack{k = 1 \\ k \neq i}}^{N}  \sum\limits_{f'=1}^{F}  \Xout_{k,f',t} \acirf \bar{P}_{k,t} H_{k,j} }  \geq \gammaT \} .  \label{computeY'}
\end{equation}
A link $(i,j)$ is said to be \textit{falsely claimed to be successful} in RB $(f,t)$, if $\Yoriginal_{i,j,f,t} = 0$ and $\Yout_{i,j,f,t} = 1$.

\begin{algorithm}[t]
	\caption{Method for Sensitivity Removal}
	\label{Alg_sensitiviyRemoval}
	\begin{algorithmic}[1]
		\Require{$\{N,F,T,\Hb,\mathbf{\acir},\bar{P}_{i,t},\gammaT,\sigma^2 \}$}
		\Ensure{$\Xb$}
		\State Compute $\bar{I}_{i,j}\,\,\, \forall\,(i,j)$ using (\ref{Iij_computation}). 
		\State Create the problem formulation [M2].
		\Do
		\State Solve the problem [M2] to find out the scheduling matrix $\Xoutb$ and successful link status matrix $\Youtb$.  \label{alg_solveProblem}
		\State Compute $\mathbf{\Yoriginal}$ using (\ref{computeY'}).
		\ForAll{$\{(\bar{i},\bar{j},\bar{f},\bar{t}): \Yout_{\bar{i},\bar{j},\bar{f},\bar{t}}=1, \Yoriginal_{\bar{i},\bar{j},\bar{f},\bar{t}} = 0\}$}
		\State Find $\Ss'$ using Algorithm \ref{Alg_computeS}.
		\State Find $\Qs$ using (\ref{Qs_computation}).
		\State Add the cover inequalities (\ref{coverInequallity2}) to [M2].
		\EndFor
		\doWhile{$\Youtb  \neq \Yoriginal$}		
	\end{algorithmic}
\end{algorithm}
\begin{algorithm}[H]
	\caption{Method for finding out $\Ss'$}
	\label{Alg_computeS}
	\begin{algorithmic}[1]
		\Require{$\{\Ss,\Hb,\mathbf{\acir},\bar{\Pb},\bar{i},\bar{j} \}$}
		\Ensure{$\Ss$}  
		\Statex // Find the minimum cardinality set $\Ss'$  \nonumber
		\State $\Ss' = \emptyset$
		\While{$\sum\limits_{(k,r) \in \Ss'}  \acir_{r} \bar{P}_{k,\bar{t}} H_{k,\bar{j}} \leq  \bar{I}_{\bar{i},\bar{j}}$}
		\State $(k',r') = \underset{(k,r) \in \Ss \setminus \Ss'}{\arg\max}   \{\acir_{r} \bar{P}_{k,\bar{t}} H_{k,\bar{j}}\}    $
		\State $\Ss' = \Ss' \cup \{(k',r')\}$
		\EndWhile
		\Statex // Maximize the RB gaps in $\Ss'$
		\While{$\sum\limits_{(k,r) \in \Ss'}  \acir_{r} \bar{P}_{k,\bar{t}} H_{k,\bar{j}} >  \bar{I}_{\bar{i},\bar{j}} $ }
		\State $(k',r') = \underset{(k,r) \in  \Ss'}{\arg\min}   \{(\acir_{r} - \acir_{r+1})\bar{P}_{k,\bar{t}} H_{k,\bar{j}}\}    $
		\State $\Ss' = (\Ss' \setminus \{(k',r')\})\cup \{(k',r'+1)\}$
		\EndWhile
		\State $\Ss' = (\Ss' \setminus \{(k',r'+1)\})\cup \{(k',r')\}$
	\end{algorithmic}
\end{algorithm}

Let us assume that in the current solution, the link $(\bar{i},\bar{j})$ scheduled in RB $(\bar{f},\bar{t})$ is a \textit{falsely claimed to be successful} link, i.e., $\Yoriginal_{\bar{i},\bar{j},\bar{f},\bar{t}} = 0$ and $\Yout_{\bar{i},\bar{j},\bar{f},\bar{t}} = 1$. Our idea is to add strong cover inequalities to the original problem, so that the link $(\bar{i},\bar{j})$ in RB $(\bar{f},\bar{t})$ will not be a \textit{falsely claimed to be successful} link in any of the future iterations with the current interference scenario. Therefore, the rest of this section focuses on generating the cover inequalities for the link $(\bar{i},\bar{j})$ in RB $(\bar{f},\bar{t})$ alone. Later on, we repeat the same procedure for generating cover inequalities for all \textit{falsely claimed to be successful} links. 

\rephrase{Let $\Ss \subseteq \Ns \times \{0,1,\cdots,F-1\}$ be the set of tuples, with each tuple containing the interferer VUE and the corresponding scheduled frequency slot gap from $\bar{f}$}, in timeslot $\bar{t}$. That is,
\begin{equation}
\Ss = \{(k,r): \exists f\in\Fs, \mathrm{s.t., } \Xout_{k,f,\bar{t}}=1, r=|f-\bar{f}|, k \neq \bar{i} \} .
\end{equation}
In other words, $\Ss$ defines an interference scenario for the link $(\bar{i},\bar{j})$ in RB $(\bar{f},\bar{t})$. Since the link $(\bar{i},\bar{j})$ in RB $(\bar{f},\bar{t})$ is a failure, we know that $\sum\limits_{(k,r) \in \Ss}  \acir_{r} \bar{P}_{k,t} H_{k,j} > \bar{I}_{\bar{i},\bar{j}}$.  We can ensure $\Yout_{\bar{i},\bar{j},\bar{f},\bar{t}}=\Yoriginal_{\bar{i},\bar{j},\bar{f},\bar{t}}$ for the same interference scenario, in all future iterations, by adding the following cover inequality to the problem formulation [M2],
\begin{multline} \label{coverInequallity1}
Y_{\bar{i},\bar{j},f,\bar{t}} \leq |\Ss| -  \sum_{(k,r) \in \Ss} V_{i,k,r,\bar{t}}    \quad \forall\, f
\end{multline}
Observe that, the right hand side of the above cover inequality is zero for the current interference scenario $\Ss$, thereby enforcing $Y_{\bar{i},\bar{j},\bar{f},\bar{t}}=0$ in the next iteration.

However, we can tighten the cover inequality (\ref{coverInequallity1}) as follows,
\begin{equation} \label{coverInequallity2}
(|\Qs| + 1) Y_{\bar{i},\bar{j},f,\bar{t}} \leq |\Ss'| + |\Qs| -  \sum_{(k,r) \in \Ss' \cup \Qs} V_{i,k,r,\bar{t}}    \quad \forall\,f 
\end{equation}
where $\Ss' \subseteq \Ss$ is the minimal cardinality set which is sufficient to cause enough interference to make the link $(\bar{i},\bar{j})$ in RB $(\bar{f},\bar{t})$ a failure, i.e., $\sum\limits_{(k,r) \in \Ss'}  \acir_{r} \bar{P}_{k,t} H_{k,j} > \bar{I}_{\bar{i},\bar{j}}$. In other words, $\Ss'$ is the list of highest interference causing elements within $\Ss$, i.e.,
\begin{equation}  
\Ss' = \underset{\Ss'' \subseteq \Ss}{\arg\min} \{|\Ss''| : \sum_{(k,r) \in \grave{\Ss}} \acir_r \bar{P}_{k,t} H_{k,j} > \bar{I}_{i,j}   \}  \label{findSs'} .
\end{equation}

\noindent To tighten the cover inequality further, we increment the RB gaps in $\Ss'$ to maximally possible values, in such a way that any further increment of any RB gaps in the resulting $\Ss'$ would result in insufficient interference to break the link $(\bar{i},\bar{j})$ in RB $(\bar{f},\bar{t})$. The computation of $\Ss'$ is explained in Algorithm \ref{Alg_computeS}. Also. we lift the cover inequality \eqref{coverInequallity2} further using the set $\Qs \subseteq \Ns \times \Fs$ \cite{Zonghao1}. The set $\Qs$ is the set of strong interferes and the corresponding RB gap tuples, which causes more interference to the link $(\bar{i},\bar{j})$  than the interference caused by any of the interferer in $\Ss$. That is,
\begin{equation} \label{Qs_computation}
\Qs = \{(k',r'):  \acir_{r'} \bar{P}_{k',t} H_{k',\bar{j}} >  \max\limits_{(k,r) \in \Ss}  \{ \acir_{r} \bar{P}_{k,t} H_{k,\bar{j}} \}   \} .
\end{equation}

Moreover, observe that, the cover inequality \eqref{coverInequallity2} can be applied for all timeslots $t$, if $\bar{P}_{i,t} = \bar{P}_{i,t'}, \,\forall\,i, t, t'$.


\section{Performance Evaluation}    \label{simulationResults}

\newcommand{\plotAextraXTickPos}{1,2,3,4,5,6,7,8,9,10,11,12,13,14,15,16,17,18,19,20}
\newcommand{\plotAxTicklabels}{2,4,6,8,10}		
\newcommand{\folderName}{Datafiles}
\newcommand{\ymaxBandC}{2.05}  \newcommand{\ymaxA}{6.2}
\newcommand{\yTicksBandC}{1,1.5,2} 


\pgfplotscreateplotcyclelist{colorList1}{
	{black,mark=square,mark options={fill=none}},
	{green,mark=triangle},
	{blue,mark=o},
	{violet,mark=star},
	{red,mark=diamond},
}

\begingroup
\thickmuskip=0mu		

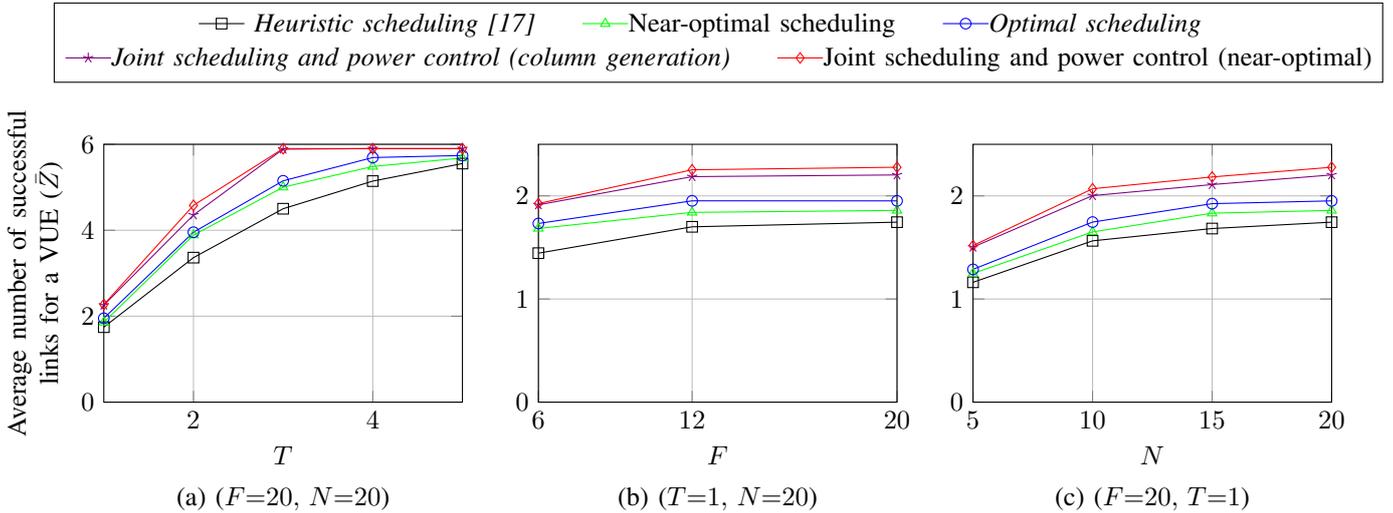
\begin{figure*}[!ht]
	\centering	

        \begin{tikzpicture}
\begin{groupplot}[cycle list name=colorList1, xmajorgrids, ymajorgrids,
	group style={group name=my plots,group size= 3 by 1,vertical sep=2.5cm },
	height=5cm,width=0.3\paperwidth	
	]

\nextgroupplot[
xmin=1, xmax=5,
ymin=0,ymax=6,
xlabel=$T$,
xtick =  \plotAxTicklabels,
]
\addplot +[restrict expr to domain={\coordindex}{0:9}]   table [x=xValues, y=sch_Heuristic1, col sep=comma] {\folderName/schPerformance_T.csv}; \label{figA:plot1} 
\addplot +[restrict expr to domain={\coordindex}{0:9}]   table [x=xValues, y=sch_gurobi, col sep=comma] {\folderName/schPerformance_T.csv}; \label{figA:plot2} 
\addplot +[restrict expr to domain={\coordindex}{0:9}]   table [x=xValues, y=sch_sensitivity1, col sep=comma] {\folderName/schPerformance_T.csv}; \label{figA:plot3} 
\addplot +[restrict expr to domain={\coordindex}{0:9}]   table [x=xValues, y=joint_diYuan1, col sep=comma] {\folderName/schPerformance_T.csv}; \label{figA:plot4} 
\addplot +[restrict expr to domain={\coordindex}{0:9}]   table [x=xValues, y=joint_basic1, col sep=comma] {\folderName/schPerformance_T.csv}; \label{figA:plot5} 

\nextgroupplot[
xmin=6, xmax=20,
ymin=0,ymax=2.5,
xlabel=$F$,
xtick =  {6,12,20},
legend style={at={(0.5,1.08)},anchor=south, legend columns=-1, column sep = 0.1cm}, 
]
\addplot  table [x=xValues, y=sch_Heuristic1, col sep=comma] {\folderName/schPerformance_F.csv}; 
\addplot  table [x=xValues, y=sch_gurobi, col sep=comma] {\folderName/schPerformance_F.csv};
\addplot  table [x=xValues, y=sch_sensitivity1, col sep=comma] {\folderName/schPerformance_F.csv};
\addplot  table [x=xValues, y=joint_diYuan1, col sep=comma] {\folderName/schPerformance_F.csv};
\addplot  table [x=xValues, y=joint_basic1, col sep=comma] {\folderName/schPerformance_F.csv};
\coordinate (top) at (rel axis cs:0,1);

\nextgroupplot[
xmin=5, xmax=20,
ymin=0,ymax=2.5,
xlabel=$N$,
]
\addplot  table [x=xValues, y=sch_Heuristic1, col sep=comma] {\folderName/schPerformance_N.csv};
\addplot  table [x=xValues, y=sch_gurobi, col sep=comma] {\folderName/schPerformance_N.csv};
\addplot  table [x=xValues, y=sch_sensitivity1, col sep=comma] {\folderName/schPerformance_N.csv};
\addplot  table [x=xValues, y=joint_diYuan1, col sep=comma] {\folderName/schPerformance_N.csv};
\addplot  table [x=xValues, y=joint_basic1, col sep=comma] {\folderName/schPerformance_N.csv};

\coordinate[right=-5.7cm] (bot) at (rel axis cs:1,0);
\end{groupplot}
\node[below = 1cm of my plots c1r1.south] {(a) ($F=20,\, N=20$)};
\node[below = 1cm of my plots c2r1.south] {(b) ($T=1,\, N=20$)};
\node[below = 1cm of my plots c3r1.south] {(c) ($F=20, \, T=1$)};

\path (top-|current bounding box.west)--
node[anchor=south,rotate=90] {\parbox{4.5cm}{\centering Average number of  successful \\ links for a VUE ($\bar{Z}$)}}
(bot-|current bounding box.west);
\path[yshift=2cm] (top|-current bounding box.north)--
coordinate(legendpos)
(bot|-current bounding box.north);
\matrix[
matrix of nodes,
anchor=south,
draw,
inner sep=0.2em,
draw
]at([yshift=1ex]legendpos)
{
	\ref{figA:plot1} \em Heuristic scheduling  \cite{AnverArchive} ~~~
	\ref{figA:plot2}\em Near-optimal scheduling ~~~
	\ref{figA:plot3}\em Optimal scheduling ~~~  \\
	\ref{figA:plot4}\em Joint scheduling and power control (column generation) ~~~
	\ref{figA:plot5}\em Joint scheduling and power control (near-optimal) \\};
\end{tikzpicture}

\caption{Average number of successful links for a VUE ($\bar{Z}$) for various scheduling algorithms} \label{figNSuccessfulLinks}
\end{figure*}

\endgroup


\pgfplotscreateplotcyclelist{colorList1}{
	{black,mark=square,mark options={fill=none}},
	{green,mark=triangle},
	{blue,mark=o},
	{violet,mark=star},
	{red,mark=diamond},
}

\begingroup
\thickmuskip=0mu		

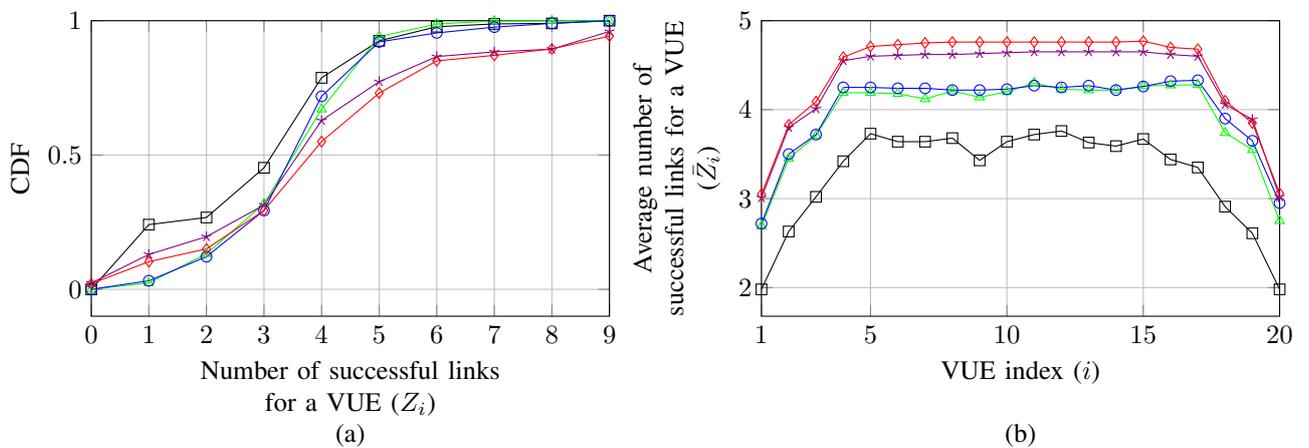
\begin{figure*}[!ht]
	\centering	

        \begin{tikzpicture}
\begin{groupplot}[cycle list name=colorList1, xmajorgrids, ymajorgrids,
	group style={group name=my plots,group size= 2 by 1,vertical sep=2.5cm ,horizontal sep=2cm },
	height=5.5cm,width=0.4\paperwidth	
	]

\nextgroupplot[
xmin = 0,
xmax = 9,
xlabel=\parbox{4cm}{\centering Number of successful links \\ for a VUE ($Z_i$) },
ylabel=CDF,
xtick =  {0,1,2,3,4,5,6,7,8,9},
ymax = 1,
]
\addplot table [x=xValues, y=sch_Heuristic1, col sep=comma] {\folderName/cdfSchPerformance.csv}; \label{figB:plot1} 
\addplot table [x=xValues, y=sch_gurobi, col sep=comma] {\folderName/cdfSchPerformance.csv}; \label{figB:plot2} 
\addplot table [x=xValues, y=sch_sensitivity1, col sep=comma] {\folderName/cdfSchPerformance.csv}; \label{figB:plot3} 
\addplot table [x=xValues, y=joint_diYuan1, col sep=comma] {\folderName/cdfSchPerformance.csv}; \label{figB:plot4} 
\addplot table [x=xValues, y=joint_basic1, col sep=comma] {\folderName/cdfSchPerformance.csv}; \label{figB:plot5} 

\nextgroupplot[
xmin = 1,
xmax = 20,
xlabel=VUE index ($i$),
ylabel=\parbox{4cm}{\centering Average number of successful links for a VUE ($\bar{Z}_i$) },
xtick =  {1,5,10,15,20},
ymax = 5,  
legend style={at={(-0.15,1.08)},anchor=south, legend columns=-1, column sep = 0.1cm}, 
]
\addplot table [x=xValues, y=sch_Heuristic1, col sep=comma] {\folderName/eachVUESchPerformance.csv};
\addplot table [x=xValues, y=sch_gurobi, col sep=comma] {\folderName/eachVUESchPerformance.csv};
\addplot table [x=xValues, y=sch_sensitivity1, col sep=comma] {\folderName/eachVUESchPerformance.csv};
\addplot table [x=xValues, y=joint_diYuan1, col sep=comma] {\folderName/eachVUESchPerformance.csv};
\addplot table [x=xValues, y=joint_basic1, col sep=comma] {\folderName/eachVUESchPerformance.csv};

%

\coordinate (bot) at (rel axis cs:1,0);
\end{groupplot}
\node[below = 1.3cm of my plots c1r1.south] {(a)};
\node[below = 1.3cm of my plots c2r1.south] {(b)};

\end{tikzpicture}

\caption{Fairness comparison for number of successful links ($F=20,\, T=2,\, N=20$)}	 \label{figFairness}
\end{figure*}

\endgroup

\subsection{Scenario and Parameters}  \label{scenarioAndParameters}

We want to emphasize that our problem formulations and algorithms do not assume any particular topology or system parameters. However, for the simulation purpose and for the ease of reproducibility, we stick with a fairly simple system model and topology. The parameters of interest are summarized in Table~\ref{table:Simulation_Parameters}. 

The topology of the vehicles consists of $N$ VUEs distributed on a convoy. The distance between any two adjacent VUEs, $d$, follows a shifted exponential distribution, with the minimum distance $d_{\text{min}}$ and the average distance $d_{\text{avg}}$. That is, the probability density function of $d$ is given as,
\begin{equation}
f(d) = \left\{ \begin{array}{ll}
(1/(d_{\text{avg}}-d_{\text{min}}))  \exp({-\frac{d-d_{\text{min}}}{d_{\text{avg}}-d_{\text{min}}}}) , & \mbox{$d \ge d_{\text{min}} $} \\
0, & \mbox{otherwise} \\
\end{array} \right. 
\end{equation}
We choose $d_{\text{avg}} = 48.6$\,m which corresponds to 2.5 seconds for a vehicular speed of $70\,$km/h, as recommended by 3GPP \cite[section A.1.2]{36.885} for freeway scenario, and we assume $d_{\text{min}} = 10\,$m . 

We adopted the channel model and channel parameters from \cite{Karedal}, which is a model based on the realtime measurements of V2V links at carrier frequency 5.2 GHz in a highway scenario. The pathloss model for a distance $d$ in \cite{Karedal} is as follows,
\begin{equation}
\text{PL}(d) = \text{PL}_0 + 10 n\hspace*{0.05cm}\text{log}_{10}(d/d_0) + X_{\sigma_1}
\end{equation}
where $n$ is the path loss exponent, $\text{PL}_0$ is the path loss at a reference distance $d_0$, and $X_{\sigma_1} $ represents the shadowing effect modeled as a zero-mean Gaussian random variable with standard deviation $\sigma_1$. An additional attenuation of $10$\,dB is added as penetration loss for each obstructing VUE \cite{Taimoor}. The noise variance is $-95.2\,$dBm and $\Pmax$ is $24\,$dBm as per 3GPP recommendations \cite{36.942}. We assume that $\gammaT = 5\,$dB  is sufficient for a transmission to be successful (i.e., the error probability averaged over the small-scale fading is sufficiently small). 

The ACIR values $\acir_{r}$ is taken same as the mask specified by 3GPP \cite{36.942}, which is as follows,
\begin{align}
\acir_{r} &= \left\{ \begin{array}{lll}
1, & \mbox{$r=0 $} \\
10^{-3}, & \mbox{$1 \leq r \leq 4 $} \\
10^{-4.5}, & \mbox{otherwise} \\
\end{array} \right.
\end{align}

For the simulation purpose, the we choose set of intended receivers $\mathcal{R}_i$ for a transmitting VUE $i$ as the closest $\min(N-1,FT-1)$ VUEs to VUE $i$ based on the distance between the VUEs.

\begin{table}[t]
	\centering
	\caption{System Simulation Parameters}
	\label{table:Simulation_Parameters}
	\begin{tabular}{|l|l|}
		\hline
		Parameter & Value \\
		\hline
		ACIR model & 3GPP mask~\cite{36.942} \\
		$\gammaT$ & $5$ dB \\
		$\Pmax$ & $24$ dBm \\
		$\text{PL}_0$ & $63.3$ dB \\
		$n$ & $1.77$  \\
		$d_0$ & 10 m \\
		$\sigma_1$ & $3.1$ dB \\
		Penetration Loss & $10$ dB per obstructing VUE \\
		$\sigma^2$ & $-95.2$ dBm\\
		\davg & 48.6 m \\
		$d_{\text{min}}$ & 10 m \\
		$\beta$ & $1/(N \Pmax)$ \\
		$\eta$ & $\gammaT(N\Pmax+\sigma^2) $ \\
		$ \Ct $ & $100$ \\
		\hline
	\end{tabular}
\end{table}

\subsection{Simulation Results} \label{subsec:SimulationResults}

To measure the performance, we use the number of successful links for a VUE, defined as, 
\begin{align}
& Z_i = \sum_{j \in \mathcal{R}_i}  Z_{i,j},   \label{defZi}\\
& \bar{Z}_i = \E[Z_i],   \label{defZibar}\\   
& \bar{Z} = \frac{1}{N}\sum_{i=1}^N  \bar{Z}_i ,  \label{defZbar}
\end{align}
where $Z_i$ is the number of successful links from VUE $i$, when VUE $i$ is transmitting a packet to all VUEs in the set $\mathcal{R}_i$. The quantity $\bar{Z}_i$ is the expected value of $Z_i$, where the expectation is taken over the random quantities in the experiment, i.e., the inter-VUE distances and shadow fading. Finally, $\bar{Z}$ is the number of successful links for a VUE, averaged across all VUEs. In other words, the metric $\bar{Z}$ can be interpreted as the average number of receiving VUEs that can decode a packet from a certain VUE. Clearly, $\bar{Z}$ must be sufficiently large to support the application in mind. However, to specify a minimum value of $\bar{Z}$ is out of scope of this paper.  

As a baseline, we also present the result of the heuristic scheduling algorithm proposed in \cite{AnverArchive} as the black curves in Figs.~\ref{figNSuccessfulLinks}--\ref{figFairness}. Due to the high numerical sensitivity of the problem, solving the scheduling problem formulation in its current form will provide only near-optimal solutions, as shown in the green curves in Figs.~\ref{figNSuccessfulLinks}--\ref{figFairness}. Therefore, we have computed optimal solution for the scheduling problem by using the sensitivity reduction techniques explained in Section \ref{sec:sensitivity}, and the results are shown in blue colored curves. Similarly, the purple colored curves indicate the performance of joint scheduling and power control using column generation method as explained in Section \ref{sec:columnGeneration}, and the red curves indicate the corresponding near-optimal performance.

We note that, when $N\leq FT$, an ACI-unaware scheduling and power control scheme is trivial, i.e., schedule all VUEs in non-overlapping RBs, and allocate maximum transmit power to each VUE. However, as illustrated in our previous work \cite{AnverArchive}, this scheme show significantly performance degradation compared to ACI-aware schemes.

In \figref{figNSuccessfulLinks}(a), we plot $\bar{Z}$ for various schemes by varying the number of timeslots $T$. As one can observe, the performance get saturated to 6 successful links for a VUE as we increase the number of timeslots $T$. This is because, the link beyond 3rd neighbor on each side of a transmitting VUE is getting noise limited due to the high penetration loss of intermediate blocking VUEs. This also implies that, if ACI can be completely avoided, each VUE can communicate upto 6 neighboring VUEs, when there are sufficient number of RBs to allocate (i.e., $FT \geq N$).
In \figref{figNSuccessfulLinks}(b), we plot $\bar{Z}$ for various values of frequency slots $F$. We note that, when $F=6$ or $12$, then multiple VUEs are getting scheduled in single RB, thereby allowing CCI. Similarly, in \figref{figNSuccessfulLinks}(c), we show the performance for various number of VUEs $N$. As we increase $N$, the performance get increases since more and more receivers are becoming available for each transmission.

In \figref{figFairness} we compare the fairness of each schemes. In \figref{figFairness}(a), we plot the CDF of the number of successful links for a VUE. Observing the steepness of the CDF of each plot, we note that optimal scheduling provides more fairness compared to the joint optimal scheduling and power control schemes. One possible explanation would be that the performance improvement for an optimal scheme is mainly due to the exploitation of asymmetry in the system. For instance, if the channel gain is same for all pairs of VUEs, then the performance improvement of optimal schemes would be marginal compared to a naive scheme. We note that, joint optimal scheduler and power controller is exploiting the asymmetry in the channel matrix more efficiently than the optimal scheduler, which may lead to the fairness degradation.

Similar fairness degradation for joint optimal scheduler and power control can also be seen in \figref{figFairness}(b). In this figure, we compare the average number of successful links for a VUE, where VUEs are indexed according to their positions in the convoy. We observe that the VUEs in the middle of the convoy are able to successfully broadcast its packet to more neighbors, since the VUEs in the middle have got more number of close-by neighbors.

\section{Conclusions}

This paper studies performance of V2V broadcast communication by focusing more upon the scenario where ACI is dominant over CCI due to the non-overlapping scheduling of VUEs. From the results presented in this paper, which are for half-duplex communication, we can draw the following conclusions,
\begin{enumerate}
	\item Performance is majorly limited by ACI when VUEs are multiplexed in frequency. However, proper scheduling and power control schemes can be used to mitigate the impact of ACI.
	\item The joint scheduling and power control problem to maximize the connectivity between VUEs, in the presence of ACI, can be modeled as an MBLP problem. From this problem formulation, we can derive scheduling alone problem as a BLP problem, and power control alone problem as MBLP.
	\item The joint scheduling and power control problem's computational complexity can be reduced using a column generation approximation method, without compromising much upon the performance.
	\item Due to the high dynamic range of channel parameters and ACI values, the problem formulation is highly numerically sensitive. This results in solver returning near-optimal solutions, instead of optimal solutions. However, optimal scheduling performance can be computed by applying proper cover inequalities to the standard problem formulation.
	\item Near-optimal joint scheduling and power control performs significantly better than optimal scheduling. However, the fairness is less for joint schemes compared to optimal scheduling schemes.
\end{enumerate}

\begin{appendices} 

\section{Computation of $V_{i,k,r,t}$} \label{Appendix:V_definition}
Let the matrix $\Ub \in \{0, 1, \cdots, F\}^{N \times T}$ indicate the scheduling of VUEs, i.e., $U_{i,t}$ is the scheduled frequency slot of VUE $i$ during the timeslot $t$. We can compute $U_{i,t}$ from $\Xb$ as follows,
\begin{equation}
U_{i,t} = \sum\limits_{f=1}^{F} f X_{i,f,t}		\quad\forall\,i,\,t  \label{compute_Uit}
\end{equation}
Observe that when VUE $i$ is not scheduled during the timeslot $t$, then $U_{i,t} = 0$. 

Next we introduce a variable $\bar{U}_{i,t} \in \{0,1\}$ which indicate if VUE $i$ is scheduled in timeslot $t$ or not. In other words,
\begin{align}
\bar{U}_{i,t}  = 
\begin{cases}
1, & \text{$U_{i,t} \neq 0$}  \\
0, & \text{Otherwise}
\end{cases}
\end{align}
We can implement the above definition using the following constraints,
\begin{subequations} \label{constr_Uit}
	\begin{align}
	\bar{U}_{i,t} &\in \{0,1\} \\
	\bar{U}_{i,t} &\geq U_{i,t}/F  \\
	\bar{U}_{i,t} &\leq U_{i,t} \label{Uit_upperBound}
	\end{align}
\end{subequations}
However, the constraint (\ref{Uit_upperBound}) is redundant, since the solver always tries to reduce $\bar{U}_{i,t}$. This is because, setting $\bar{U}_{i,t} = 1$ results in reduction of feasible region as we will see from the following paragraphs. 

Now, we can define $V_{i,k,r,t} \in \{0,1\}$ which indicates if VUE $i$ and $k$ are scheduled in RBs which are not more than $r$ RBs far apart in timeslot $t$. That is,
\begin{equation} \label{define:Vikrt}
V_{i,k,r,t}  = 
\begin{cases}
1, & \text{If $\bar{U}_{i,t} = 1, \bar{U}_{k,t} = 1, |U_{i,t} - U_{k,t}| \leq r $}  \\
0, & \text{Otherwise}
\end{cases}
\end{equation}
To mathematically translate the above definition into a set of linear constraints, we introduce two Boolean auxiliary variables $V'_{i,k,r,t}$ and $V''_{i,k,r,t}$, as follows,
\begin{subequations}
	\begin{align}
	V'_{i,k,r,t}  &= 
	\begin{cases}
	1, & \text{If $U_{i,t} - U_{k,t} \leq r$}  \\
	0, & \text{Otherwise}
	\end{cases}    \label{def:Vikrt'} \\
	V''_{i,k,r,t}  &= 
	\begin{cases}
	1, & \text{If $U_{k,t} - U_{i,t} \leq r$}   \label{def:Vikrt''} \\
	0, & \text{Otherwise}
	\end{cases} 
	\end{align}	
\end{subequations}
Now we can implement the definition (\ref{define:Vikrt}), using the auxiliary variables $V'$ and $V''$ as follows,
\begin{subequations} \label{constr_Viklt}
	\begin{align}
	&V_{i,k,r,t} \leq \bar{U}_{i,t} \\
	&V_{i,k,r,t} \leq \bar{U}_{k,t} \\
	&U_{i,t} - U_{k,t} \leq r + \eta'(1-V'_{i,k,r,t})   \\
	&U_{i,t} - U_{k,t} \geq r + 1  -\eta' V'_{i,k,r,t}   \\
	&U_{k,t} - U_{i,t} \leq r + \eta'(1-V''_{i,k,r,t})	 \\
	&U_{k,t} - U_{i,t} \geq r + 1  -\eta' V''_{i,k,r,t}   \\
	&V_{i,k,r,t} \geq V'_{i,k,r,t} + V''_{i,k,r,t} - 1 	\\
	&V_{i,k,r,t} \leq V'_{i,k,r,t} 	\\
	&V_{i,k,r,t} \leq V''_{i,k,r,t}  \\ 	
	&V_{i,k,r,t}, V'_{i,k,r,t}, V''_{i,k,r,t} \in \{0,1\}
	\end{align}
\end{subequations}
The constraints (\ref{constr_Viklt}a--b) ensure that $V_{i,k,r,t}$ can be 1 only if both VUEs $i$ and $k$ are scheduled. 
The constraints (\ref{constr_Viklt}c--d) implements the definition (\ref{def:Vikrt'}), i.e., ensure that $V'_{i,k,r,t} = 1$, if and only if $U_{i,t} - U_{k,t} \leq r$. Similarly, the constraints (\ref{constr_Viklt}e--f) implements the definition (\ref{def:Vikrt''}). The parameter $\eta'$ is a sufficiently large number which makes the constraints (\ref{constr_Viklt}c--f) redundant for the appropriate values of $V'_{i,k,r,t}$ and $V''_{i,k,r,t}$. It is not hard to show that $\eta'=2F$ is sufficient. 
The constraints (\ref{constr_Viklt}g--i) ensures $V_{i,k,r,t} = V'_{i,k,r,t} \land V''_{i,k,r,t}$, where $\land$ represents AND operation of two boolean variables. Also, observe that the constraints (\ref{constr_Viklt}h--i) are redundant, since the solver is trying to reduce $V_{i,k,r,t}$, therefore, an upper bound is not required for $V_{i,k,r,t}$.

\end{appendices}

\bibliographystyle{IEEEtran}
\bibliography{jointSchedulingAndPowerControl}	   
\end{document}